# Twisted-light-revealed Lightlike Exciton Dispersion in Monolayer MoS$_2$

Kristan Bryan Simbulan[1,§], Teng-De Huang[1,§], Guan-Hao Peng[2§], Feng Li[3], Oscar Javier Gomez Sanchez[2], Jhen-Dong Lin[2], Junjie Qi[3], Shun-Jen Cheng[2,*], Ting-Hua Lu[1,*] and Yann-Wen Lan[1,*]

[1]*Department of Physics, National Taiwan Normal University, Taipei 11677, Taiwan*
[2]*Department of Electrophysics, National Chiao Tung University, Hsinchu 30010, Taiwan*
[3]*School of Materials Science and Engineering, University of Science and Technology Beijing, Beijing 100083, People's Republic of China*

\* Corresponding authors
§ Authors contributed equally to this work

**Abstract:** Twisted light carries a well-defined orbital angular momentum (OAM) per photon. The quantum number $\ell$ of its OAM can be arbitrarily set, making it an excellent light source to realize high-dimensional quantum entanglement and ultra-wide bandwidth optical communication structures. To develop solid-state optoelectronic systems compatible to such promising light sources, a timely challenging task is to efficiently and coherently transfer the optical OAM of light to certain solid-state optoelectronic materials. Among the state-of-the-art emergent materials, atomically thin monolayer transition metal dichalcogenide (ML-TMD), featured by ultra-strong light-matter interaction due to its reduced dimensionality, renders itself a potential material suitable for novel applications. In this study, we carried out photoluminescence (PL) spectroscopy studies of ML-MoS$_2$ under photoexcitation of twisted light with well-defined quantized OAM. We mainly observed pronounced increases in the spectral peak energy for every increment of $\ell$ of the incident twisted light. The observed non-linear $\ell$-dependence of the spectral blue shifts evidences the OAM transfer from the exciting twisted light to the valley excitons in ML-TMDs, which is well accounted for by our analysis and computational simulation. Even more excitingly, the twisted light excitation is shown to make excitonic transitions relative to the transferred OAM, enabling us to infer the exciton band dispersion from the measured spectral shifts. Consequently, the measured non-linear $\ell$-dependent spectral shifts revealed an unusual lightlike exciton band dispersion of valley excitons in ML-TMDs that is predicted by previous theoretical studies and evidenced for the first time via our experimental setup that utilizes the unique twisted light source.

## I. Introduction

Optical Vortex beam (VB), also called twisted light, is a stream of light with a phase singularity and carries a well-defined orbital angular momentum (OAM)[1] equivalent to $\ell\hbar$ per photon. Its helicoidal wavefront gives rise to a topological structure uniquely identified by the quantum number of its OAM $\ell$. Twisted light have been considered in various research fields leading to advanced discoveries and applications[2] in, but not limited to, communication systems[3], optical manipulation[4,5], quantum entanglement[6,7], enhanced imaging systems[8,9], chemistry[10] and biomedicine[11]. Due to its unique properties, a handful of researchers have also become interested in exploring the helical light beam's interaction with condensed matter. Twisted light-matter interaction (TLMI) involves the transfer of light beam's OAM onto illuminated objects[12–14] in such a way that the total angular momentum is conserved. A new degree of freedom (DOF) can be introduced by twisted light through its quantum number $\ell$ that can be exploited to uncover certain phenomena that are challenging to observe in an experimental setting. Such is revealed by Schmiegelow *et al.* (2016) when they demonstrated enhanced electric quadrupole transitions in a trapped $^{40}Ca^+$ ion that leads to a modified optical selection rule[14]. Opportunities to expand existing applications may also arise, such as the work of Sordillo *et al.* (2019) in which findings suggest significance in spintronics[15]. Through the introduction of OAM to the exciting light, it is possible that certain predicted properties in other materials may be tested or new behaviors may be discovered. One of these prospective materials with very interesting properties are the layered transition metal dichalcogenides (TMDs).[16] Interlayer quasiparticles in TMDs are recently reported to have exhibited quasi-angular momentum – a new DOF – via a near-zero twist angle heterostructure[17]. Excitingly, the twisted wavefront of an incident twisted light beam may also induce a distinct DOF in the TMD material by virtue of its properties.

Photo-exciting a ML-TMD with twisted light should give rise to new exciton transitions that has acquired additional finite momenta from the transferred OAMs of light[18,19]. With every increment of the quantized OAM of the incident twisted light, the photo-induced exciton transitions in a ML-TMD might shift towards high momentum regime and, correspondingly, yield blue-shifted PL energies. For such scenario, our optical spectroscopic facility set up with a twisted light source serves like an alternative tool that can sample the exciton energies in the distinct momenta regime and allows one to infer the exciton band dispersion which is so far still not measurable. The $k$-dispersion of the exciton band of a photo-excited material is enriched with exciton physics[20–22]. Because of the inherently weak Coulomb screening in low dimensional materials, excitons in photo-excited ML-TMDs are subjected to extraordinarily strong Coulomb interaction, including both the electron-hole direct and exchange interactions. It is the former that makes an exciton bound so tightly in a 2D material, while the latter is known to be the main cause of fast valley depolarization of excitons – one of the most relevant physics in the application of valleytronics where long valley times are essentially necessary and desired[23–29]. From previous theoretical studies, the underlying electron-hole exchange interaction in an exciton leads to an unusual light-like exciton dispersion[20,21]. In this work, we show that our optical measurement with the unique twisted light source evidences for the first time the exchange-interaction-driven light-like exciton band of ML-TMDs. Specifically, we detected meV-scale increases in the measured PL peak spectral energy as the $|\ell|$ of the incident twisted light was incremented. The observed non-linear $\ell$-dependence pattern of the detected PL spectral shifts is shown to be a signature of the predicted linear lightlike exciton band dispersion, which is well accounted for by our simulation of the twisted light and monolayer MoS$_2$ interaction.

II. **Methods**

To perform Raman characterization and PL spectral measurements, a micro-Raman spectroscopy setup was used as shown in **Fig. 1(a)**. Inserted along the optical path prior to the sample is a spatial light modulator (SLM) responsible for the generation of the twisted light. Here, we use a set of computer-generated holograms to carry out spatial modulation to a 532 nm continuous laser. Each hologram is characterized by a particular phase corresponding to a value of $\ell$ that will be imposed on the incident beam. The reflected beam would then have an OAM of $\ell\hbar$ per photon. **Fig. 1(b)** illustrates the phase, wavefront and the intensity profiles associated with selected values of $\ell$. The twisted light from the SLM was tightly-focused on the sample using a 100x objective lens. The setup was designed so that the power and the $\ell$ of twisted light can be varied before an optical measurement was executed. All measurements were done at room temperature as excitonic effects in ML and bilayer (BL) TMDs remain strong within this operating condition[30].

Triangular MoS$_2$ samples were prepared by chemical vapor deposition on a Si/SiO$_2$ substrate. Deliberately grown onto the substrate were stacked layers of triangular MoS$_2$ among which a specific sample, as seen in **Fig. 2(a)** and **2(b)**, was selected. Raman spectrum of the outer triangular portion of the sample (**Fig.**

**2(c))** shows a peak separation of 19.2 cm$^{-1}$ between the two dominant phonon modes of MoS$_2$, which indicates a ML structure[31]. Meanwhile, the Raman spectra of its inner triangular part (**Fig. 2(d)**) shows a peak separation of 21.9 cm$^{-1}$, which closely corresponds to a BL configuration[32].

### III. Results and Discussion

At a fixed laser power of 800 μW, we performed repeated PL measurements each at different values of $\ell$ of the incident twisted light on both the ML- and the BL-MoS$_2$ samples. Clear shifts of the A exciton PL peaks from long wavelength (at $\ell = 0$) to shorter wavelengths (at $1 \leq |\ell| \leq 4$) were observed in the ML-MoS$_2$ sample (**Fig. 3(a) and Fig. S1(b) to S1(c)**), which corresponds to an increase in the photoemission energy (blue shift) as the $|\ell|$ of the incident light increases (**Fig. 3(c)**). It is also apparent that these spectral blue shifts follow a non-linear $\ell$-dependence pattern. The overall change in energy of the ML-MoS$_2$ sample's PL peaks, from when $\ell = 0$ to $\ell = \pm 4$, is about 2.6 meV (spectral shift of 0.97 nm). Note that the small spectral shifts recorded in between measurements is not a measurement deviation as they are well above the spectrometer's resolution (0.05 nm). In fact, to prove the abovementioned claim, a set of relatively larger PL spectral shifts from another triangular ML-MoS$_2$ sample (**Fig. S2**) showed a similar pattern as that of **Fig. 3(a) and 3(c)** indicating that both sets of data, regardless of the difference in the magnitudes of their respective spectral shifts, have possibly manifested the same phenomenon.

To verify the involvement of the properties of the exciting twisted light in this observed phenomenon, we conduct a theoretical analysis of the light-matter interactions (LMI), as well as a numerical simulation of the optical spectra, of a ML-MoS$_2$ under the excitation of twisted light as follows.

**Twisted Light and ML-MoS$_2$ interaction**

In the above experiments, the PL spectra of the ML-MoS$_2$ were obtained by exciting the sample *non-resonantly* and by subsequently detecting the spontaneously emitted photons from the excitons therein, which, prior to the exciton-photon conversion, undergo some energy relaxation processes. For simplicity of modelling, we assume resonant excitation and simply evaluate the rate of the generic optical transitions (light emission or absorption) disregarding the processes of relaxation. Note that the PL spectral shifts induced via resonant excitation is also measured and shown in **Fig. S3**, which is very similar to that of the non-resonant case (**Fig. 3(a) and 3(c)**) and supportive of the adopted assumption. According to the Fermi's golden rule, the rate of the generic optical transition of a 2D material under the weak excitation of light with vector potential $\vec{A}(\vec{r},t)$ and angular frequency $\omega$ is evaluated by

$$w_{S\vec{Q}}(\omega) = \frac{2\pi}{\hbar}\left|\left\langle \Psi_{S\vec{Q}}^X \middle| \widehat{H}_I \middle| GS \right\rangle\right|^2 \delta\left(E_{S\vec{Q}}^X - \hbar\omega\right), \qquad (1)$$

where $\widehat{H}_I$ is the second quantized operator of the leading linear term in the light-matter interaction, $H_I \equiv \frac{|e|}{2m_0}\vec{A}(\vec{r},t)\cdot\vec{p}$, in the Coulomb gauge and in the rotating wave approximation[33], e ($m_0$) is the elementary charge (rest mass) of electron, $\vec{p}$ is the linear momentum, $|GS\rangle$ is the ground state of the sample under no excitation, $|\Psi_{S\vec{Q}}^X\rangle$ denotes the single exciton state in the exciton band labelled by the band index $S$ and the center-of-mass momentum of exciton $\vec{Q}$, and $E_{S\vec{Q}}^X$ is the energy of the exciton state.

For ML-MoS$_2$, the lowest optically active exciton states belong to the two lowest like-spin exciton bands composed of the K and K' valley exciton states that, in the small-$\vec{Q}$ regime, can be well modeled by the effective pseudo-spin Hamiltonian[20,21]. In the effective pseudo-spin model, the K and K' valley exciton states form a spinor and the inter-valley electron-hole exchange interaction in the valley exciton acts as an effective transverse field that splits the bands of the K- and K'-valley exciton at finite $\vec{Q}$ (see Supporting Information for the review). Accordingly, we calculate the exciton band dispersions of a ML-MoS$_2$ in the exciton model with the use of the density-functional-theory (DFT)-calculated parameters of the exciton's effective masses and transition dipoles following the methodology in Ref.[22], as presented in **Fig. 4(a)**. After solving the eigenstates of the valley exciton in the pseudo-spin model, the dipole moment of the exciton in the $S_\pm$-band with small $\vec{Q}$ is obtained as $\vec{D}_{S_\pm\vec{Q}}^X = \sqrt{\Omega}\, D_0^X\left(\hat{\varepsilon}_+ \pm e^{i2\varphi_{\vec{Q}}}\hat{\varepsilon}_-\right)$ with $D_0^X \approx 0.128e$, where $\Omega$ is the area of the 2D material, $\varphi_{\vec{Q}}$ is the azimuthal angle of the wave vector $\vec{Q} = Q(\cos\varphi_{\vec{Q}}, \sin\varphi_{\vec{Q}})$, and $\hat{\varepsilon}_\pm = (\hat{x} \pm i\hat{y})/\sqrt{2}$ is the unit vector of the $\sigma_\pm$-polarized dipole moments of the K/K' valley exciton (See Supporting Information for the detailed information). As known from previous studies[20,21], the $\vec{Q}$-dependent electron-hole exchange interaction splits the valley-exciton bands into a *parabolic* lower band $\left(E_{S_-\vec{Q}}^X\right)$ and a nearly *linear* upper one $\left(E_{S_+\vec{Q}}^X\right)$. Analytically, it can be shown that $E_{S_+\vec{Q}}^X = 2\gamma Q$ and $E_{S_-\vec{Q}}^X = \frac{\hbar^2}{2M_{\text{eff}}^X}Q^2$, where $\gamma \equiv (D_0^X)^2/2\epsilon_0 = 1.47(\text{eV}\cdot\text{Å})$ and $M_{\text{eff}}^X \sim 1.08m_0$ are determined by the DFT-based computations as presented in Ref.[22]. The linearity of the upper band, $E^X_{S_+\vec{Q}}$, with a linear dispersion manifests itself as a direct consequence of the electron-hole exchange interaction, known to be essential in the valley dynamics of excitons[27,29,34–40]. By contrast, the lower band remains to be in the parabolic and nearly flat dispersion, featured by a very narrow bandwidth (~ μeV) as a consequence of the cancellation of the intra- and inter-valley electron-hole exchange interactions and the Coulomb interaction enhanced exciton mass[21]. **Fig.4(a)** shows the bright exciton bands of ML-MoS$_2$ around the light cone area, which are split by the electron-hole exchange interaction at the scale of few meV. The observed blue shifts of the spectral peaks in our experiments are typically about or above a few meV in some samples, implying the involvement of the linear upper exciton band in the optical spectroscopy.

Let us now turn to the modelling of the light-matter interaction. In the Lorenz gauge[41,42], the vector potential of the LG beam is written as

$$\vec{A}_{\hat{\varepsilon}\ell p q_0}(\vec{r},t) = \hat{\varepsilon}\, A_{\ell p}(\vec{r}_\parallel) \exp[i(q_0 z - \omega t)] \qquad (2)$$

where $\hat{\varepsilon}$ is the unit vector of light polarization, $q_0$ is the magnitude of the wave vector along the propagating (z-) direction, $\vec{r}_\parallel = (\rho, \varphi)$ is the in-plane position in polar coordinates, $A_{\ell p}(\vec{r}_\parallel)$ is the amplitude of the LG beam spatially varying in the plane normal to the z-direction[41,42], and the subscript $\ell$ (p) denotes the quantum number of OAM (radial mode index) of the LG beam. The validity of the use of the vector potential of **Eq. (2)** in Lorenz gauge in the light-matter interaction of **Eq. (1)**, which is yet in Coulomb gauge, is justified in Ref. [43]. For the simplicity of analysis, we set p = 0 throughout this work and take the compact form of the LG beam's amplitude, that is $A_\ell(\vec{r}_\parallel) = \frac{A_0}{\sqrt{|\ell|!\pi/2}} \left(\frac{\sqrt{2}\rho}{W_0}\right)^{|\ell|} \exp(-\rho^2/W_0^2)\exp(i\ell\varphi)$, where $A_0$ is the constant of amplitude and $W_0$ is the beam waist of the LG beam. To better identify the interaction of the twisted light with an exciton possessing the in-plane center-of-mass momentum in a 2D material, we decompose the spatially varying amplitude of the twisted light into a series of plane waves with in-plane wave vectors $\vec{q}_\parallel$,

$$A_\ell(\vec{r}_\parallel) = \sum_{\vec{q}_\parallel} \mathcal{A}_\ell(\vec{q}_\parallel) \exp[i\vec{q}_\parallel \cdot \vec{r}_\parallel]. \qquad (3)$$

After some algebra, the amplitude of the $\vec{q}_\parallel$ - component is solved as $\mathcal{A}_\ell(\vec{q}_\parallel) = \left(\frac{\pi A_0 W_0^2 (-i)^\ell}{\Omega\sqrt{|\ell|!\pi/2}}\right)\left[\left(\frac{q_\parallel W_0}{\sqrt{2}}\right)^{|\ell|} \exp\left(-\frac{q_\parallel^2 W_0^2}{4}\right)\right] e^{i\ell\varphi_{\vec{q}_\parallel}}$ (See Supporting Information for the detailed derivation).

Assuming the sufficiently broad bandwidth of the exciting light and taking the electric dipole approximation, the total rate of the optical transition to create or annihilate an exciton in the state $\Psi_{S\vec{Q}}^X$ is obtained by integrating $w_{S\vec{Q}}^{\hat{\varepsilon}\ell}(\omega)$ over all $\omega$,

$$W_{S\vec{Q}}^{\hat{\varepsilon}\ell} = \int d\omega \, w_{S\vec{Q}}^{\hat{\varepsilon}\ell}(\omega) \propto \left|\mathcal{A}_\ell(\vec{Q})(\hat{\varepsilon}\cdot \vec{D}_{S\vec{Q}}^{X\,*})\right|^2, \quad (4)$$

which can be formulated in terms of the product of the $\vec{Q}$-component of the amplitude of the LG beam, $\mathcal{A}_\ell(\vec{Q})$, and the projection of the exciton's transition dipole moment $\vec{D}_{S\vec{Q}}^X$ onto the direction of $\hat{\varepsilon}$ (See Supporting Information for details). **Fig. 4(b)** plots the function $\left|\mathcal{A}_\ell(\vec{Q})\right|^2$ versus $\vec{Q}$, which accounts for the distribution of the optically accessible exciton states in the $\vec{Q}$-space by the $\ell$-mode of the exciting LG beam. As one can observe, the function $\left|\mathcal{A}_\ell(\vec{Q})\right|^2$ is shifted towards higher $\vec{Q}$-regime as the $\ell$ of the exciting twisted light increases. This indicates that the twisted light with higher $|\ell|$ optically excites more exciton states with greater Q and higher energies.

To simulate the spectral shift of the optical peak caused by the twisted light, we specify a spectral function of energy, $g_S^\ell(E)$, which resolves the total transition rate $W_S^{\hat{\varepsilon}\ell} \equiv \sum_{\vec{Q}} W_{S\vec{Q}}^{\hat{\varepsilon}\ell} = \frac{\Omega}{4\pi^2}\int d^2\vec{Q}\, W_{S\vec{Q}}^{\hat{\varepsilon}\ell}$ in the energy spectrum, as implicitly defined by

$$\frac{\Omega}{4\pi^2}\int d^2\vec{Q}\, W_{S\vec{Q}}^{\hat{\varepsilon}\ell} = \int dE\, g_S^\ell(E). \quad (5)$$

Since our experiments were carried out at room temperature, thermal distribution is so broad that the spectral function $g_S^\ell(E)$ alone, defined as the distribution function of the transition rate in energy, should well model the profile of the PL spectral peak. To find out $g_S^\ell(E)$, we carry out the integral of the left-hand side of **Eq. (5)** and transform the variable Q to E via the predicted band dispersion of exciton, $E_{S_\pm\vec{Q}}^X$ (See Supporting Information for the details).

**Numerical Simulation of the spectral shifts of the twisted light-induced optical peaks**

**Fig. 4(c)** presents the spectral function $g^\ell(E) = g^\ell_{S_-}(E) + g^\ell_{S_+}(E)$ with respect to $\ell$, which, in agreement with the experimental results, shifts towards higher energy as $\ell$ is increased. Correspondingly, **Fig. 4(d)** shows the spectral shift as a function of $\ell$ defined by $\Delta E_\ell \equiv \langle E \rangle_\ell - \langle E \rangle_{\ell=0}$, where $\langle E \rangle_\ell \equiv \frac{\int dE E g^\ell(E)}{\int dE\, ^\ell(E)}$ is the averaged energy of the spectral function $g^\ell(E)$. Taking into account the dielectric screening of ML-MoS$_2$ on the Si/SiO$_2$ substrate, the simulated energy blue shifts induced by the twisted light are scaled up by the factor of the effective refractive index, $n_{MoS_2} \sim 4.8^{44}$, and turn to fall in a similar scale as that of the observed shift (~few meV). Notably, one can see that the simulated spectral energy shift $\Delta E_\ell$ follows a non-linear $\ell$-dependence (**Fig. 4(d)** in blue dots) – a feature very similar to the measured $\ell$-dependence of the PL energy as shown in **Fig. 3(c)**. Analytically, one can derive the closed form of $\Delta E_\ell$ as a function of $\ell$ and show that the non-linear $\ell$-dependence of $\Delta E_\ell$ is associated with the dispersion linearity of the dominant upper exciton band, as the manifestation of the electron-hole exchange interactions that plays the key role in the dynamics of valley polarization of exciton. For more illustration, we purposely ignore the electron-hole exchange interaction in the model calculation to retain the dispersion parabolicity of the exciton bands (See the inset of **Fig. 4(a)**). We show that the parabolicity of the exciton band dispersion yet leads to the linear $\ell$-dependence of $\Delta E_\ell$ (See the red dots in **Fig. 4(d)**), as predicted by our analysis. More detailed analysis regarding the association of the spectral shift's non-linear $\ell$-dependence with the linear exciton dispersion is given in Sec. III of the Supporting Information, and concluded by Eqs.(S37) and (S39).

While our model calculation qualitatively well predicts the non-linear $\ell$-dependent feature of the experimentally-observed blue shift, very few quantitative discrepancies still exist between the theoretical and the experimental results. The predicted energy shift falling in the scale of one meV is close to, but smaller than, the measured values by few meVs. Some causes of the discrepancies are speculated. Firstly, the exciton transition in the current theory is modeled in the electric dipole approximation. It has been reported that the quadrupole interaction should also play a certain role in the light-matter interaction under the twisted light excitation[45]. The effects of the quadrupole interaction definitely deserve further studies and remain as an open subject for future research. Secondly, rigorously speaking, the accurate refractive index of the atomically thin ML-TMDs, which reduces the wavelength of the exciting light inside the sample and affects the energies of the light-accessible exciton states, is not surely known. The existing theories for the determination of the effective refractive indices of atomically thin 2D materials mostly follow the theory of macroscopic optics[44,46,47] rather than the microscopic theory suited for nano-materials[48]. Lastly, it may also be necessary to consider the effects of the $\ell$-dependent power density of the incident twisted light on the sample. Theoretically, the higher the $\ell$ of the incident light, the lower its power density on the surface of the sample. The reason is that, as the $\ell$ of the twisted light increases, its beam spot expands leading to a higher effective area. This can, thus, raise a possibility that the laser's power density could also influence the observed PL measurements. **Fig. S4** shows the typical amount of red shift caused by the incident fundamental mode light's power density on the ML-$MoS_2$ sample, where an increase of power from 0.1 mW to 0.8 mW results in a 0.0824 meV per kW/cm$^2$ power density increase. In comparison, the blue shift observed on the ML-$MoS_2$ in **Fig. 3(a)** and **3(c)** is relatively higher at 0.897 meV for every kW/cm$^2$ increase in power density. Thus, it can be claimed that, while the effect of the twisted light is dominant in the observed results, the difference between the theoretical and the measured spectral shifts could be partially due to the effect of the incident beam's $\ell$-dependent power density.

Measured PL spectra taken from the BL-$MoS_2$ sample (**Fig. 3(b)** and **Fig. S1(e)** to **S1(f)**), on the other hand, shows similar behavior as in ML-$MoS_2$ such that clear non-linear $\ell$-dependent blue shifts (**Fig. 3d**) of the A exciton peaks with increasing |$\ell$| were also recorded. Compared to ML, BL-$MoS_2$ exhibits a maximum overall shift of 17.38 meV ($\ell = 0 \rightarrow -4$). The larger shifts in BL-$MoS_2$ is speculated as the consequence of a possibly steeper exciton band dispersion resulting from the interplay between the electron-hole exchange interaction and inter-layer coupling in the material. The LMI mechanism, however, in the BL-$MoS_2$ case is not guaranteed to be the same with that of the ML-$MoS_2$ as various interactions may have also occurred in the former. Further studies regarding this matter is still necessary.

Power-dependence of the PL spectra from the ML-$MoS_2$ sample illuminated with twisted light is also investigated with the results shown in **Fig. 5**. Successive PL measurements after each increment of laser power, with a fixed value of $\ell$, were done to complete one cycle over a total of four cycles of measurements. Each cycle is performed using different values of $\ell$ of the incident light, that is at $0 \leq \ell \leq 3$ in incrementing order. Based on the results, a clear dependence of the photoemitted energy to increasing laser power was observed. This is an expected behavior as it was also seen in a previous report[49]. Here, thermal effects due to laser heating or photo-induced current[50] may become more significant, which lead to a red shift in the PL peak as the power of the incident light increases. This demonstration reveals the sensitivity of the $MoS_2$ material, not only to the quantum number $\ell$ of the twisted light, but also to the instantaneous incident laser power. Moreover, the recorded data at higher value of $\ell$, when the sample has already been exposed to the laser at a relatively longer time, might have already experienced the heating effect which is evident by the relatively lower blue shift compared to the one presented in **Figure 3a**.

The above discussions were only focused on the analysis of A exciton peaks taken from both ML- and BL-MoS$_2$ samples; the PL data concerning B excitons, specifically for BL MoS$_2$, were attached in **Fig. S5**, which shows a similar blue shift behavior. To verify the reproducibility and repeatability of all the reported results, another sample was selected and shown in **Fig. S6(a)** in the Supporting Information; similar results were obtained and shown in **Fig. S6(b) – S6(c)**.

## IV.     Conclusion

This work reported intriguing optical properties of valley excitons in ML-MoS$_2$ as a consequence of its interaction with twisted light. A noticeable blue shift was observed as the magnitude of the quantum number of OAM $|\ell|$ of the twisted light was incremented, which was well interpreted as the consequence of OAM transfer from the incident twisted light by our theoretical analysis and numerical simulation. On top of that, the study convincingly revealed, for the first time, the signature of the predicted lightlike exciton dispersion of the valley exciton in 2D MoS$_2$ materials. The peak spectral energy shift with $\ell$ that was detected in ML-MoS$_2$ was found to occur in BL-MoS$_2$ as well. The puzzling dramatic shifts in BL-MoS$_2$ are possibly related to the exciton band structures likely complicated further by the unclear interplay between the electron-hole exchange interaction and inter-layer coupling in the material, and remains as an intriguing subject for future research.

**Acknowledgment**

This work was supported by the National Science Council, Taiwan under contract No. MOST 105-2112-M-003-016-MY3, MOST 106-2112-M-009-015-MY3, and MOST 107-2112-M-003-012. This work was also in part supported by the National Nano Device Laboratories and National Center for High-performance Computing of Taiwan.

**Author contributions**

K. B. Simbulan, T.D. Huang, and G.H. Peng have contributed equally in this work. The experiments were performed by K. B. Simbulan, T.D. Huang, and F. Li. The numerical calculations were provided by G.H. Peng O.J.G. Sanchez, and J.D. Lin. J. Qi, S.J. Cheng, Y.W. Lan and T.H. Lu supervised this research. All authors have read and approved the manuscript. All authors discussed the results and commented on the manuscript.

**Competing interests**

The authors declare no competing financial interests.

**Materials and Correspondence**

Correspondence and material requests should be addressed to Y. W. Lan and T. H. Lu.

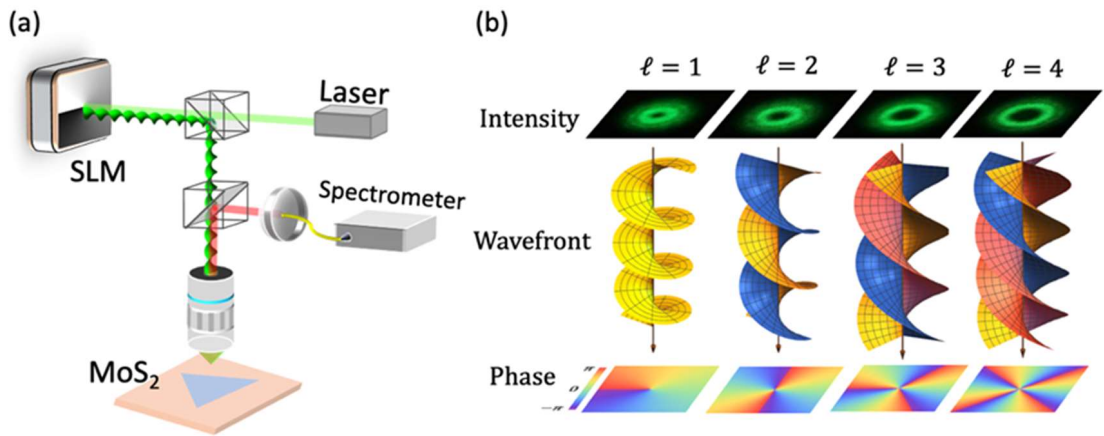

**Figure 1. (a)** Micro-Raman spectrometer setup with spatial light modulator. The spatial light modulator modulates the phase of the optical wavefront of incident light using holograms; **(b)** the intensity profile, wavefront, and phase corresponding to quantum number $\ell$ of twisted light.

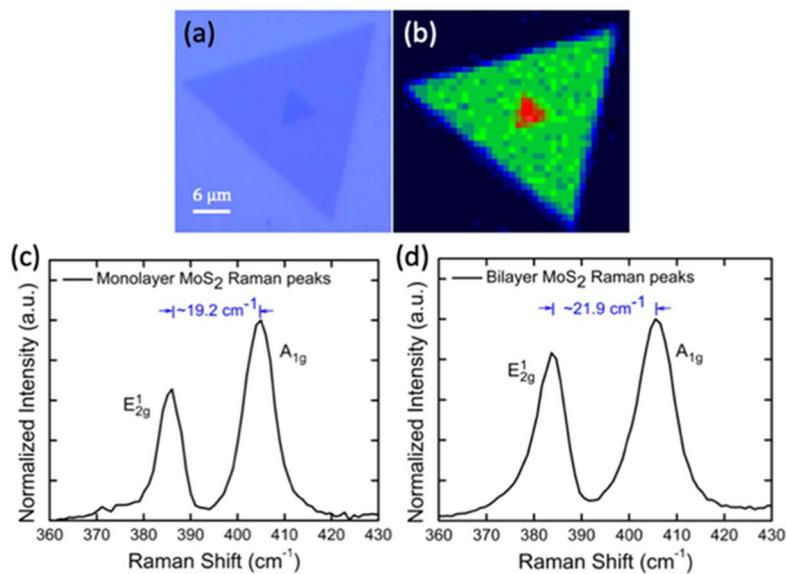

**Figure 2. (a)** Optical microscope and **(b)** Raman mapping image of the $MoS_2$ sample with AB stacking. The Raman spectra of both the **(c)** monolayer and **(d)** bilayer portion of the $MoS_2$ sample.

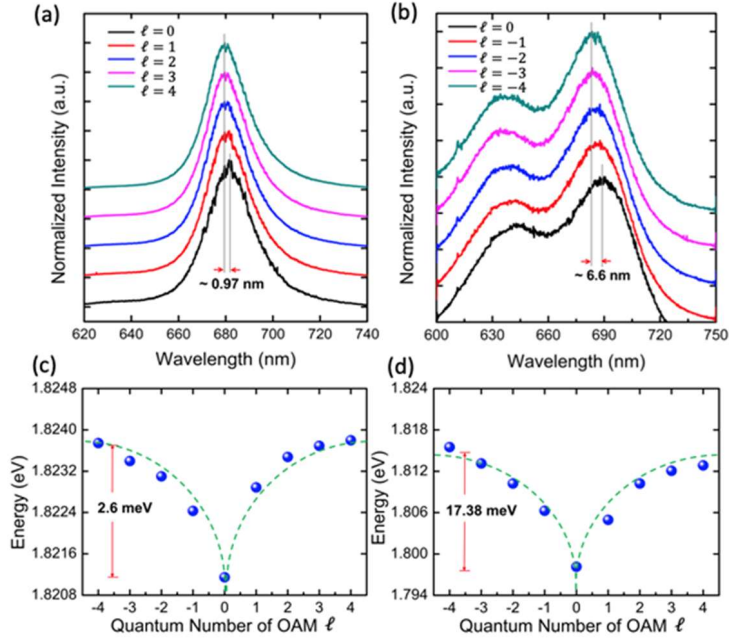

**Figure 3.** The PL spectra of the (a) monolayer and the (b) bilayer MoS$_2$ samples, using twisted light of varying values of $\ell$. Blue shifts are observed in both spectra as $|\ell|$ increases. The energy versus $\ell$ graph of the A excitons of both (c) monolayer and (d) bilayer samples are also shown for a clearer view of the energy shifts.

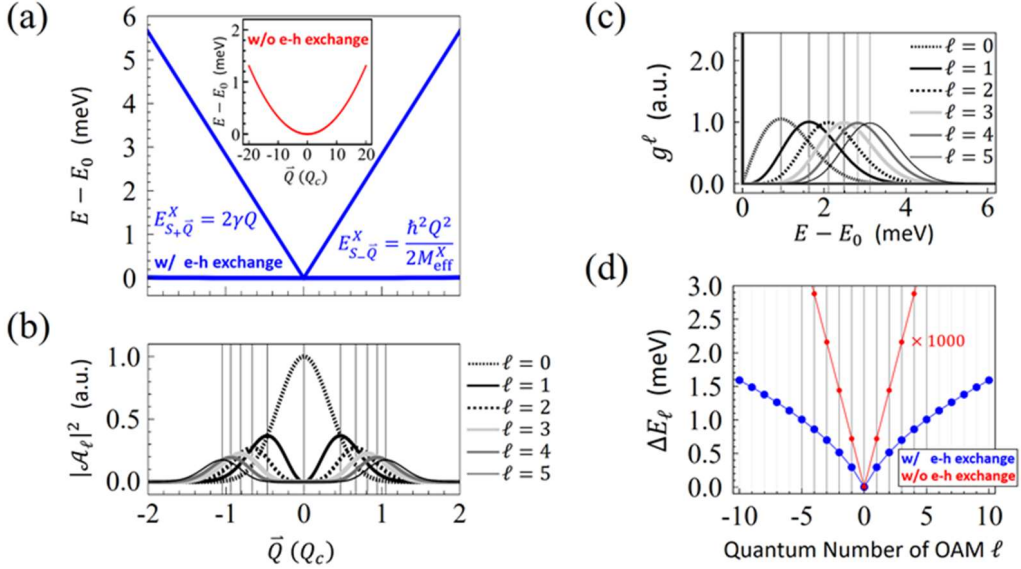

**Figure 4.** (a) The bright exciton bands of ML-MoS$_2$ (blue color) simulated using the pseudo-spin exciton model with the use of the DFT-based parameters of exciton mass and transition dipole. Due to the electron-hole exchange interaction, the upper exciton band exhibits the unusual lightlike linear dispersion, described by $E^X_{S+\vec{Q}} = 2\gamma Q$ where $\gamma = 1.47 (\text{eV} \cdot \text{Å})$. The lower exciton band dispersion described by $E^X_{S-\vec{Q}} = \frac{\hbar^2}{2M^X_{\text{eff}}} Q^2$ remains parabolic but, with the heavy exciton mass $M^X_{\text{eff}} = 1.08 m_0$, looks very flat in the small Q light cone area. $Q_c$ denotes the radius of the light cone and $E_0$ is the band edge energy of the exciton bands. The inset shows the two-fold degenerate parabolic exciton bands of ML-MoS$_2$ calculated with the neglect of the electron-hole exchange interaction (red color). (b) The magnitude of the $Q$-component in the spatially varying amplitude, $|\mathcal{A}_\ell|^2$, as a function of $Q$, of the twisted light with quantum number of OAM, $\ell$. (c) The spectral functions $g^\ell(E)$ as functions of E for $\ell = 0, 1, ..5$. (d) The spectral energy shift $\Delta E_\ell$ of ML-MoS$_2$ induced by the photo-excitation of the OAM lights with various $\ell$. The blue (red) dots show the non-linearly (linear) $\ell$-dependent spectral shifts based on the valley-split (degenerate) exciton band structure with (without) the electron-hole exchange interaction as presented in the main panel (inset) of (a).

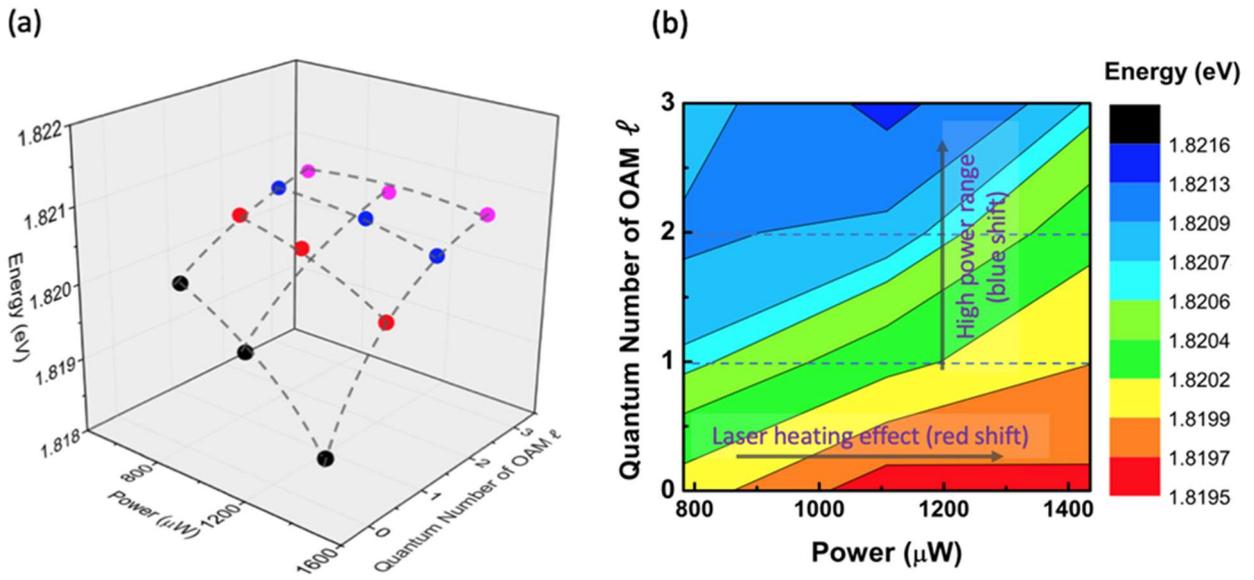

**Figure 5. (a)** A 3D diagram showing the power-dependence of the PL spectra from the monolayer sample illuminated with twisted light at different $\ell$; and the corresponding **(b)** contour chart. The dashed lines are guides to the eye.

# Supporting Information

# Twisted-light-revealed Lightlike Exciton Dispersion in Monolayer MoS$_2$

Kristan Bryan Simbulan[1,§], Teng-De Huang[1,§], Guan-Hao Peng[2§], Feng Li[3], Oscar Javier Gomez Sanchez[2], Jhen-Dong Lin[2], Junjie Qi[3], Shun-Jen Cheng[2]*, Ting-Hua Lu[1,*] and Yann-Wen Lan[1,*]

[1]*Department of Physics, National Taiwan Normal University, Taipei 11677, Taiwan*
[2]*Department of Electrophysics, National Chiao Tung University, Hsinchu 30010, Taiwan*
[3]*School of Materials Science and Engineering, University of Science and Technology Beijing, Beijing 100083, People's Republic of China*

\* Corresponding authors
§ Authors contributed equally to this work

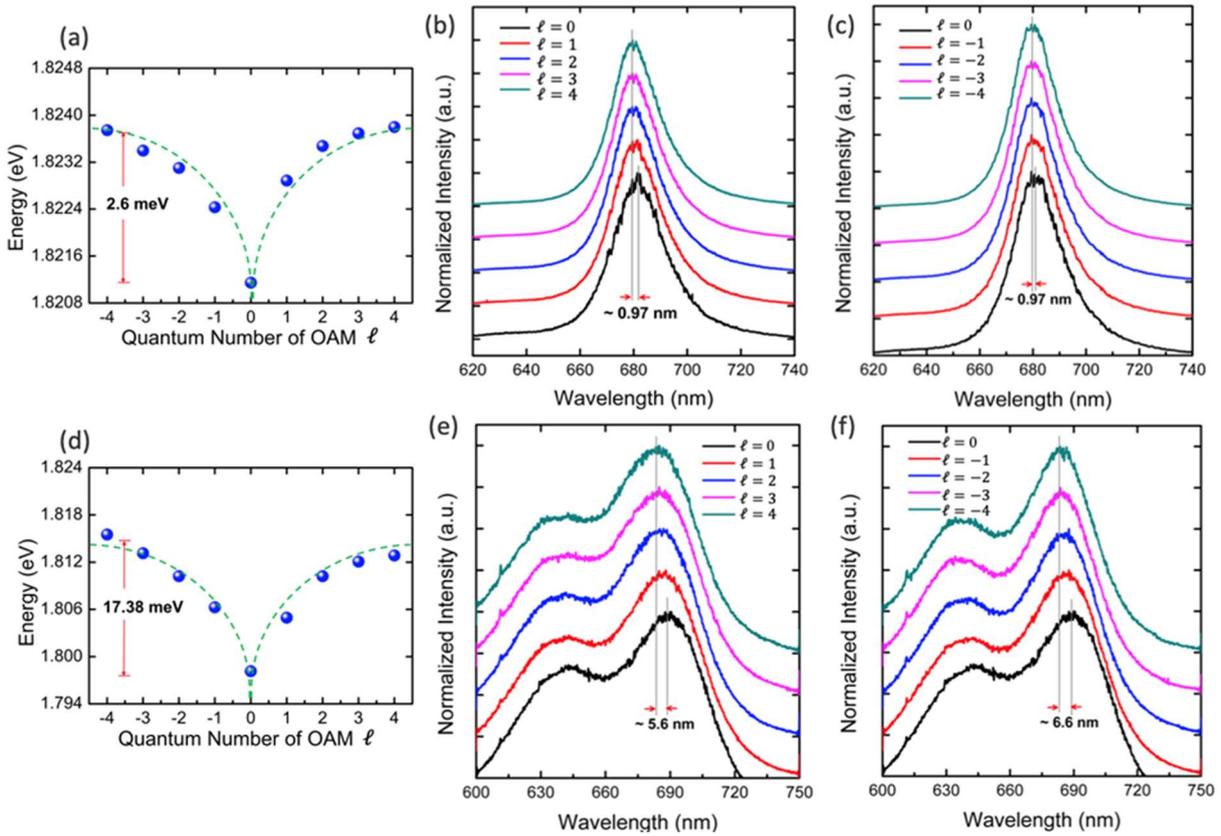

**Figure S1.** (a) The energy versus $\ell$ graph and the (b-c) corresponding PL spectra showing the energy shift of the A exciton PL peaks of the ML-MoS$_2$ sample, from the main text, that was excited with twisted light at different values of $\ell$. As $|\ell|$ increases, the photon energy increases (blue shift). Similarly, the (d) energy versus $\ell$ graph as well as the (e-f) corresponding PL spectra of the BL-MoS$_2$ sample, from the main text, showing the blue shift of its A excitons is also presented.

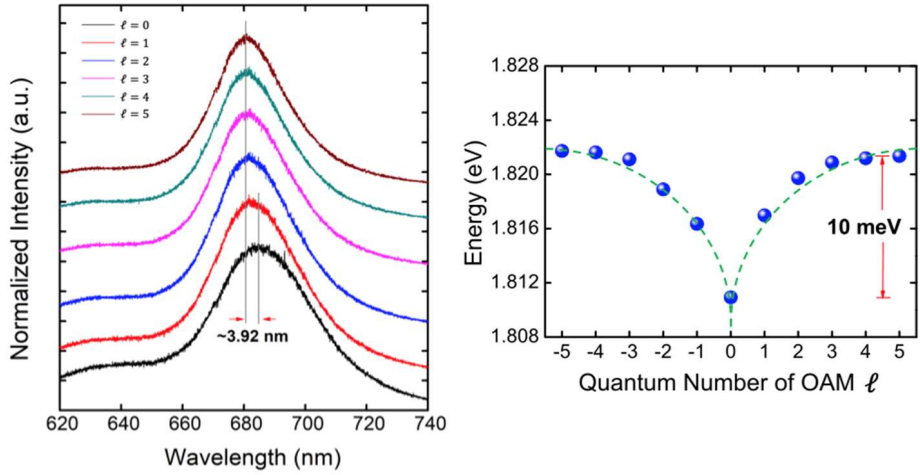

**Figure S2. Higher Energy Spectral Shift from another MoS$_2$ sample.** (a) The PL spectra showing the energy shift of the A exciton PL peaks of another monolayer MoS$_2$ sample that was excited with twisted light at different values of $\ell$; and (b) the corresponding energy versus $\ell$ graph.

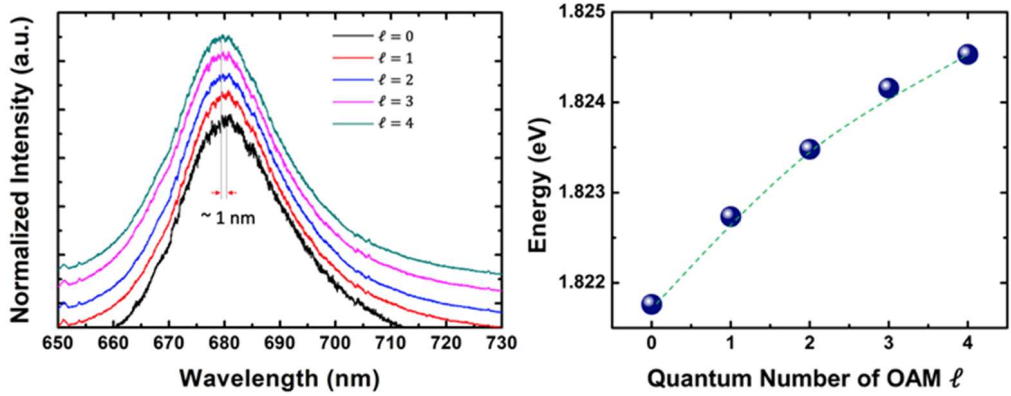

**Figure S3.** Spectral shifts of the ML-MoS$_2$ sample induced by an incident red (633 nm) laser at different $\ell$. (a) PL spectra and (b) Energy versus $\ell$ graph.

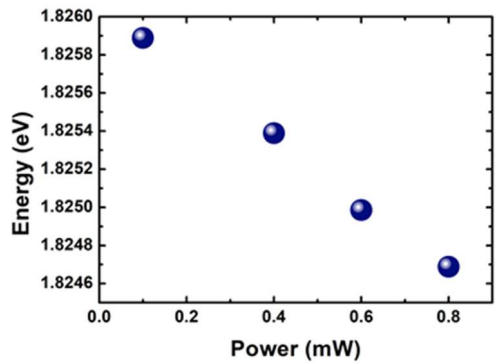

**Figure S4. Laser Thermal Heating Effect.** Red shift in the PL spectra induced by the increasing fundamental mode laser power with an effective area of 4.8x10$^{-12}$ m$^2$ on the surface of a ML-MoS$_2$ material.

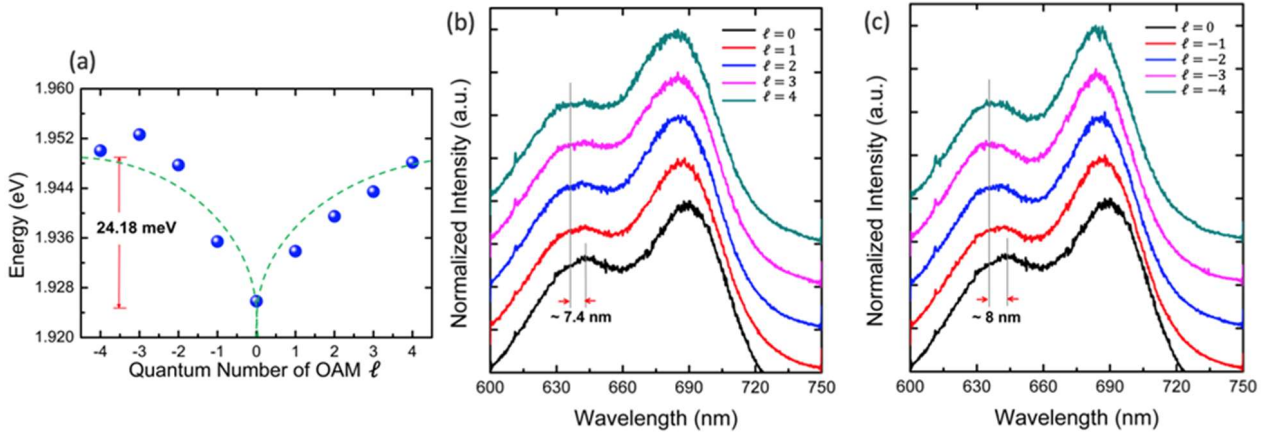

**Figure S5.** (a) The energy versus $\ell$ graph and the (b-c) corresponding PL spectra showing the energy shift of the B exciton PL peaks of the BL-MoS$_2$ sample, from the main text, that was excited with twisted light at different values of $\ell$. As $|\ell|$ increases, the photon energy increases.

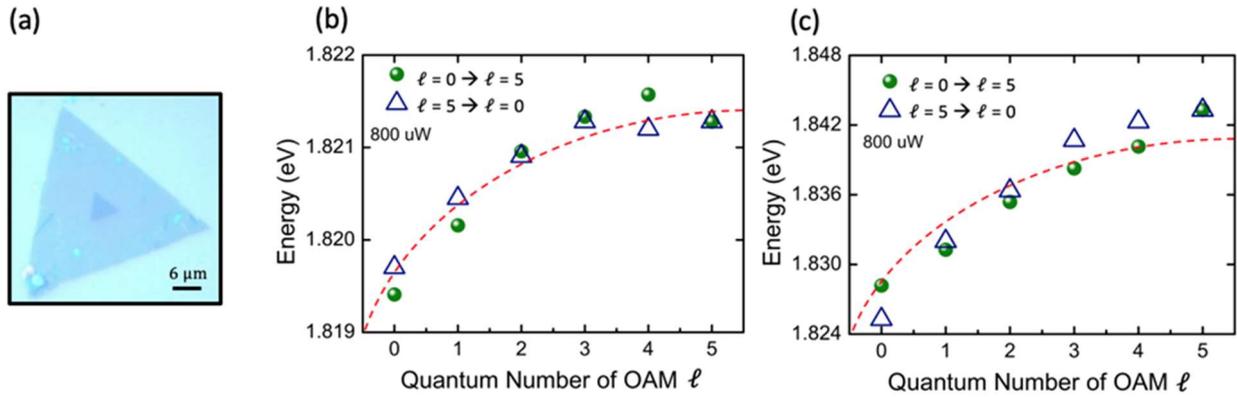

**Figure S6. Reproducible blue shift.** The (a) optical micrograph of another sample. The corresponding blue shifts, with increasing $\ell$ of the incident light, observed from the A exciton PL peaks of both the (b) monolayer and (c) bilayer MoS$_2$ portions of the sample in (a) at 800 μW laser power. Here, the value of $\ell$ is incremented first from 0 to 5 then from 5 back to 0 to show the shifts' dependence on the value of $\ell$.

## I. Theory of exciton in 2D materials

### I-1. The effective pseudospin Hamiltonian of exciton

Based on the quasi-particle Bloch state $\psi_{n\vec{k}}$ with band index $n$ and wavevector $\vec{k}$, the exciton state, $\left|\Psi^X_{S\vec{Q}}\right\rangle$, with the center-of-mass momentum $\vec{Q}$ can be written, in the second quantization language, as

$$\left|\Psi^X_{S\vec{Q}}\right\rangle = \frac{1}{\sqrt{\Omega}} \sum_{v,c} \sum_{\vec{k}} \Lambda^{vc}_{S\vec{Q}}(\vec{k}) \hat{c}^\dagger_{c\vec{k}+\vec{Q}} \hat{c}_{v\vec{k}} |GS\rangle, \tag{S1}$$

where $\Omega$ is the area of the 2D material, $S$ is the index of exciton band, $\Lambda^{vc}_{S\vec{Q}}(\vec{k})$ is the amplitude of an electron-hole configuration in the exciton wavefunction, $c(v)$ denotes the conduction (valence) band, $\hat{c}^\dagger(\hat{c})$ is the creation(annihilation) operator, and $|GS\rangle$ is the ground state of the material under no excitation. The energy and the wave function of the exciton states $|\Psi^X_{S\vec{Q}}\rangle$ can be obtained by numerically solving the density-functional theory (DFT) based Bethe-Salpeter equation as presented by Ref.[3]. Following Refs[1,2], we adopt the numerically calculated doubly degenerate bright exciton states with the disregard of the electron-hole exchange interaction as exciton basis, in which the full exciton Hamiltonian (containing both direct and exchange interactions) is reformulated to be the effective pseudo-spin Hamiltonian in the form of 2 by 2 matrix. In the effective pseudo-spin model (with the DFT-based parameters), the exciton band structures are solvable and allow for further analysis.

For MoS$_2$-MLs, the lowest optically active exciton states belong to the two lowest like-spin exciton bands that, in the small-$\vec{Q}$ regime, can be well modeled by the effective pseudo-spin Hamiltonian as presented in Refs.[1-3]. In the effective pseudo-spin model, the K and K' valley exciton states form a spinor and the inherent electron-hole exchange interaction in the valley exciton acts as an effective transverse field that split the bands of K- and K'-valley exciton at finite $\vec{Q}$. It is known that the lowest bright exciton states in photo-excited TMD-MLs are the intra-valley spin-like exciton states, $\Psi^X_{K\vec{Q}}$ and $\Psi^X_{K'\vec{Q}'}$, where the electron and hole with the same electron spins are resident in the same $K$ or $K'$ valley. The two exciton bands would be nearly degenerate if the electron-hole exchange interactions were absent. In reality, due to the inherent electron-hole exchange interactions, the nearly degenerate bright exciton bands at finite $\vec{Q}$ are actually split by $\sim$ meV and the true eigen states of exciton turn out to be the intermixed states of $\Psi^X_{K\vec{Q}}$ and $\Psi^X_{K'\vec{Q}'}$. Thus, one can take the orthogonal K and K' valley exciton states with the same $\vec{Q}$, $\Psi^X_{K\vec{Q}}$ and $\Psi^X_{K'\vec{Q}}$, as basis, in which the exciton Hamiltonian is formulated as the $2 \times 2$ pseudo-spin Hamiltonian,

$$H^X_{\text{eff}}(\vec{Q}) = \begin{pmatrix} E_0(\vec{Q}) + V^x_{K,K}(\vec{Q}) & V^x_{K,K'}(\vec{Q}) \\ V^x_{K',K}(\vec{Q}) & E_0(\vec{Q}) + V^x_{K',K'}(\vec{Q}) \end{pmatrix}, \tag{S2}$$

where $E_0(\vec{Q}) = E_0 + \hbar^2 Q^2 / 2M^X_{\text{eff}}$ describes, in terms of the exciton band edge energy $E_0$ and exciton effective mass $M^X_{\text{eff}}$, the doubly degenerate like-spin bright exciton bands in the parabolic dispersion subjected to *no* electron-hole exchange interaction, and the terms $V^x_{S,S'}(\vec{Q})$ are the matrix elements arising from the intra($S = S'$)- or inter($S \neq S'$)-valley electron-hole exchange interaction. Following the theory of Ref.[3], the matrix elements of the electron-hole exchange interaction can be formulated as

$$V^x_{S,S'}(\vec{Q}) = \frac{1}{2\epsilon_0 Q} \frac{1}{\Omega} \left[\vec{Q} \cdot \vec{D}^{X*}_{S\vec{Q}}\right] \left[\vec{Q} \cdot \vec{D}^X_{S'\vec{Q}}\right], \tag{S3}$$

where $\epsilon_0$ is the vacuum permittivity, and $\vec{D}^X_{S\vec{Q}}$ is the dipole moments of the exciton basis states $\Psi^X_{S\vec{Q}}$ that is given by

$$\vec{D}^X_{S\vec{Q}} = \frac{1}{\sqrt{\Omega}} \sum_{vc} \sum_{\vec{k}} \Lambda^{vc}_{S\vec{Q}}(\vec{k}) \vec{d}_{v\vec{k},c\vec{k}}. \tag{S4}$$

Following the methodology employed in Ref.[3], we compute $E_0(\vec{Q})$ and $\Lambda^{vc}_{S\vec{Q}}(\vec{k})$ by solving the DFT-based Bethe-Salpeter equation (BSE) and calculate $\vec{d}_{v\vec{k},c\vec{k}} \equiv e\langle\psi_{v\vec{k}}|\vec{r}|\psi_{c\vec{k}}\rangle$ in the way presented in Ref.[3]. Accordingly, we obtain $\vec{D}^X_{K\vec{Q}}/\sqrt{\Omega} \approx D^X_0(\hat{x} + i\hat{y})$ and $\vec{D}^X_{K'\vec{Q}}/\sqrt{\Omega} \approx D^X_0(\hat{x} - i\hat{y})$ with $D^X_0 \approx 0.128e$, and the exciton effective mass $M^X_{\text{eff}} = 1.08m_0$. Further, one can solve and re-formulate the intra- and inter-valley electron-hole exchange interaction of Eq.(S3), as given by $V^x_{K,K}(\vec{Q}) = V^x_{K',K'}(\vec{Q}) = \gamma Q$ and $V^x_{K,K'}(\vec{Q}) = V^{x*}_{K',K}(\vec{Q}) = \gamma Q e^{-i2\varphi_{\vec{Q}}}$ where $\gamma \equiv (D^X_0)^2/2\epsilon_0 = 1.47 (\text{eV} \cdot \text{Å})$ and $\varphi_{\vec{Q}}$ is the azimuthal angle of $\vec{Q} = Q(\cos\varphi_{\vec{Q}}, \sin\varphi_{\vec{Q}})$.

**I-2. The exciton band dispersions and the density of states**

The eigen-energies of the effective exciton Hamiltonian of Eq.(S2) are solved as

$$E^X_{S_+\vec{Q}} \approx E_0 + 2\gamma Q \tag{S5}$$

and

$$E^X_{S_-\vec{Q}} = E_0 + \hbar^2 Q^2/2M^X_{\text{eff}}, \tag{S6}$$

which belong to the upper and lower bands, respectively. As previously revealed by Refs.[1,2], the upper band exhibits the linear dispersion band while the lower band remain the parabolicity in the dispersion. Correspondingly, the eigen states of the upper linear band are explicitly given by

$$\left|\Psi^X_{S_+\vec{Q}}\right\rangle = \left(\left|\Psi^X_{K\vec{Q}}\right\rangle + e^{i2\varphi_{\vec{Q}}}\left|\Psi^X_{K'\vec{Q}}\right\rangle\right)/\sqrt{2}, \tag{S7}$$

and those of the lower band are

$$\left|\Psi^X_{S_-\vec{Q}}\right\rangle = \left(\left|\Psi^X_{K\vec{Q}}\right\rangle - e^{i2\varphi_{\vec{Q}}}\left|\Psi^X_{K'\vec{Q}}\right\rangle\right)/\sqrt{2}. \tag{S8}$$

Accordingly, the density of states (DOS) of the exciton bands defined by $\rho_S(E) \equiv \frac{\Omega}{(2\pi)^2}\int d^2\vec{Q}\,\delta\left(E^X_{S\vec{Q}} - E\right)$, are solved as

$$\rho_{S_+}(E) = \frac{\Omega}{(2\pi)^2}\int d^2\vec{Q}\,\delta(E_0 + 2\gamma Q - E) = \frac{\Omega}{2\pi}\frac{E - E_0}{(2\gamma)^2} \tag{S9}$$

and the lower parabolic band as

$$\rho_{S_-}(E) = \frac{\Omega}{(2\pi)^2}\int d^2\vec{Q}\,\delta(E_0 + \hbar^2 Q^2/2M^X_{\text{eff}} - E) = \frac{\Omega}{2\pi}\frac{M^X_{\text{eff}}}{\hbar^2}. \tag{S10}$$

Notably, the DOS of the upper linear band shows the linear proportionality to the increased energy, whereas that of the lower parabolic one remain constant and unchanged by varying the energy.

**I-3. The momentum matrix elements and transition dipole moment of valley-mixed exciton**

The momentum matrix element of a single electron transition defined as $\langle\psi_{v\vec{k}}|\vec{p}|\psi_{c\vec{k}}\rangle$ is associated with the transition dipole moment via the relationship $\vec{d}_{v\vec{k},c\vec{k}} = \frac{e\hbar}{im_0}\frac{1}{[\epsilon_{v\vec{k}} - \epsilon_{c\vec{k}}]}\langle\psi_{v\vec{k}}|\vec{p}|\psi_{c\vec{k}}\rangle$, where $\epsilon_{n\vec{k}}$ is the energy of the quasi-particle state $\psi_{n\vec{k}}$ and $m_0$ is the free electron mass (see the details in Ref.[3]). Disregarding the electron-hole exchange interaction, the momentum matrix element of the valley exciton state being a superposition of the correlated single electron transitions as expressed by Eq.(S1) is given by $\frac{1}{\sqrt{\Omega}}\sum_{vc}\sum_{\vec{k}}\Lambda^{vc}_{S\vec{Q}}(\vec{k})\langle\psi_{v\vec{k}}|\vec{p}|\psi_{c\vec{k}}\rangle \approx \frac{im_0 E_g}{|e|\hbar}\vec{D}^X_{S\vec{Q}}$, where we have used the property that the transition energy $\epsilon_{c\vec{k}} - \epsilon_{v\vec{k}}$ is nearly equal to the quasi-particle energy gap $E_g$ around the $\vec{k}$-region where $\Lambda^{vc}_{S\vec{Q}}(\vec{k})$ is not vanished. For TMD-MLs, the dipole moment

of the K (K') valley exciton is known to be $\sigma^+(\sigma^-)$-polarized and can be re-written as $\vec{D}^X_{K(K')\vec{Q}}/\sqrt{2} = \sqrt{\Omega}\, D_0^X \hat{\varepsilon}_{+(-)}$ where

$$\hat{\varepsilon}_\pm \equiv (\hat{x} \pm i\hat{y})/\sqrt{2}$$

is the unit vector of $\sigma^+(\sigma^-)$-polarization and $D_0^X$ is evaluated as $0.128e$ according to Eq.(S4).

Under the action of the electron-hole exchange interaction, the exciton eigen states turn out to be the mixture of the two distinctive valley exciton, as previously shown in Eq.(S7) and Eq.(S8). Accordingly, the dipole moments of the valley-mixed exciton states $\Psi^X_{S_\pm \vec{Q}}$ are given by

$$\vec{D}^X_{S_\pm \vec{Q}} = \frac{1}{\sqrt{2}}\left[\vec{D}^X_{K\vec{Q}} \pm e^{i2\varphi_{\vec{Q}}}\vec{D}^X_{K'\vec{Q}}\right] = \sqrt{\Omega}\, D_0^X [\hat{\varepsilon}_+ \pm e^{i2\varphi_{\vec{Q}}}\hat{\varepsilon}_-]. \quad (S11)$$

## II. Theory of orbital angular momentum light

### II-1. The vector potential of the Laguerre-Gaussian orbital angular momentum light

The vector potential of a Laguerre-Gaussian (LG) mode of orbital angular momentum (OAM) light in the Lorenz gauge is expressed as in the form,

$$\vec{A}_{\hat{\varepsilon} \ell p q_0}(\vec{r}, t) = \hat{\varepsilon}\, A_{\ell p}(\vec{r}_\parallel) \exp[i(q_0 z - \omega t)], \quad (S12)$$

where $\hat{\varepsilon}$ is the unit vector of light polarization, $\ell = 0,1,2,\ldots$ ($p = 0,1,2,\ldots$) denotes the quantum number of OAM (radial mode index) of the LG-OAM light, $q_0$ is the magnitude of the wavevector along the propagating (z-) direction, $\omega$ is the angular frequency of the light, $A_{\ell p}(\vec{r}_\parallel)$ is the amplitude of the LG-OAM light spatially varying in the plane normal to the z-direction. Following Refs.[4,5], the spatially varying amplitude of the LG-OAM light is explicitly given by

$$A_{\ell p}(\vec{r}_\parallel) = A_0 \sqrt{\frac{2(p!)}{(|\ell|+p)!\pi}}\, L_p^{|\ell|}\left(\frac{2\rho^2}{W_0^2}\right) \left(\frac{\sqrt{2}\rho}{W_0}\right)^{|\ell|} \exp\left(-\frac{\rho^2}{W_0^2}\right) \exp(i\ell\varphi), \quad (S13)$$

where $L_p^{|\ell|}(x) = \sum_{m=0}^p (-1)^m \left[\frac{(|\ell|+p)!}{(p-m)!(|\ell|+m)!m!}\right] x^m$ is the generalized Laguerre polynomial, $A_0$ is the constant of amplitude and $W_0$ is referred to as the beam waist of the OAM light., $\vec{r}_\parallel = (\rho, \varphi)$ is the position vector in the polar coordinate system and it is defined in the 2D-plane normal to the propagating direction (z-direction here). According to our experimental setups, we will use $W_0 = 1.5\mu m$ in our numerical simulations.

**II-2 The vector potential of the fundamental LG mode with $p = 0$**

For the simplicity of analysis, let us focus on the fundamental LG mode with $p = 0$ throughout this work. Taking that $L_0^{|\ell|}(x) = 1$ for all $\ell$ in Eq.(S13), one derives the amplitude of the vector potential of the LG mode with $p = 0$ in the simple form,

$$A_{\ell 0}(\vec{r}_\parallel) \equiv A_\ell(\vec{r}_\parallel) = \frac{A_0}{\sqrt{|\ell|!\,\pi/2}} \left(\frac{\sqrt{2}\rho}{W_0}\right)^{|\ell|} \exp\left(-\frac{\rho^2}{W_0^2}\right) \exp(i\ell\varphi). \tag{S14}$$

**II-3 The amplitude of LG modes expanded by in-plane plane waves**

The spatially varying amplitude of the LG-OAM light with $p = 0$ in Eq.(S14) can be expanded in terms of in-plane plane waves with the in-plane wave vector $\vec{q}_\parallel$ as

$$A_\ell(\vec{r}_\parallel) = \sum_{\vec{q}_\parallel} \mathcal{A}_\ell(\vec{q}_\parallel) \exp[i\vec{q}_\parallel \cdot \vec{r}_\parallel], \tag{S15}$$

where after some algebra the expansion coefficient $\mathcal{A}_\ell(\vec{q}_\parallel)$ is solved as

$$\mathcal{A}_\ell(\vec{q}_\parallel) = \left(\frac{\pi A_0 W_0^2 (-i)^\ell}{\Omega\sqrt{|\ell|!\,\pi/2}}\right) \left[\left(\frac{q_\parallel W_0}{\sqrt{2}}\right)^{|\ell|} \exp\left(-\frac{q_\parallel^2 W_0^2}{4}\right)\right] e^{i\ell\varphi_{\vec{q}_\parallel}}. \tag{S16}$$

In a dielectric medium, the wave length of the light is scaled down by the factor of the refractive index of the dielectric medium, $n$. In the case, Eq.(S16) should be modified by replacing the in-plane wave vector $q_\parallel$ therein by $q_\parallel/n$.

**III. The interaction between an exciton and a LG-OAM light**

The full Hamiltonian of light-matter interaction (LMI) for a material under a weak excitation of light is given by

$$\widetilde{H}_I \approx \frac{|e|}{2m_0} \vec{A}(\vec{r},t) \cdot \vec{p} + c.c. = H_I + c.c., \tag{S17}$$

where $\vec{A}(\vec{r},t)$ is the vector potential of the exciting light in the Coulomb gauge, $e$ ($m_0$) is the elementary charge (mass) of an electron, and $\vec{p}$ is the linear momentum, and

$$H_I = \frac{|e|}{2m_0} \vec{A}(\vec{r},t) \cdot \vec{p}\,. \tag{S18}$$

From the Fermi's golden rule, the rate of the optical transition that creates or annihilates an exciton in the state $\Psi_{S\vec{Q}}^X$ under the excitation of the light with the angular frequency $\omega$ is evaluated by using the formalism,

$$w_{S\vec{Q}}^{\hat{\varepsilon}\ell}(\omega) = \frac{2\pi}{\hbar} \left| \left\langle \Psi_{S\vec{Q}}^X \middle| \widehat{H}_I \middle| GS \right\rangle \right|^2 \delta\left(E_{S\vec{Q}}^X - \hbar\omega\right), \tag{S19}$$

in the rotating wave approximation, and the Hamiltonian of Eq.(S18) is expressed as $\widehat{H}_I = \sum_{n,n'} \sum_{\vec{k},\vec{k}'} \langle \psi_{n\vec{k}} | H_I | \psi_{n'\vec{k}'} \rangle \hat{c}_{n\vec{k}}^\dagger \hat{c}_{n'\vec{k}'}$ in the language of second quantization. Taking the exciton wave function in Eq.(S1), the matrix element in Eq.(S19) reads

$$\left\langle \Psi_{S\vec{Q}}^X \middle| \widehat{H}_I \middle| GS \right\rangle = \frac{1}{\sqrt{\Omega}} \sum_{v,c} \sum_{\vec{k}} \Lambda_{S\vec{Q}}^{vc*}(\vec{k}) \left\langle \psi_{c\vec{k}+\vec{Q}} \middle| H_I \middle| \psi_{v\vec{k}} \right\rangle. \tag{S20}$$

From Eq.(S12), (S15) and (S18), the matrix element of the LMI on the right hand side of Eq.(S20) driven by an OAM light reads

$$\left\langle \psi_{c\vec{k}+\vec{Q}} \middle| H_I \middle| \psi_{v\vec{k}} \right\rangle = \frac{|e|}{2m_0} \sum_{\vec{q}_\parallel} \mathcal{A}_\ell(\vec{q}_\parallel) \left[ \hat{\varepsilon} \cdot \left\langle \psi_{c\vec{k}+\vec{Q}} \middle| e^{i\vec{q}\cdot\vec{r}} \vec{p} \middle| \psi_{v\vec{k}} \right\rangle \right] e^{-i\omega t}. \tag{S21}$$

Taking the use of the property that the thickness of the 2D-material is much smaller than the wave length of the visible light, we can get

$$\left\langle \psi_{c\vec{k}+\vec{Q}} \middle| e^{i\vec{q}\cdot\vec{r}} \vec{p} \middle| \psi_{v\vec{k}} \right\rangle \approx \delta_{\vec{q}_\parallel,\vec{Q}} \left[ \hbar\vec{k} \langle u_{c\vec{k}+\vec{Q}} | u_{v\vec{k}} \rangle + \langle u_{c\vec{k}+\vec{Q}} | \vec{p} | u_{v\vec{k}} \rangle \right]. \tag{S22}$$

Furthermore, if we do the series expansion to the periodic part of the Bloch function $u_{c\vec{k}+\vec{Q}}$ at $\vec{Q}$ and take only the dominant term in the series, we can get $\hbar\vec{k} \langle u_{c\vec{k}+\vec{Q}} | u_{v\vec{k}} \rangle + \langle u_{c\vec{k}+\vec{Q}} | \vec{p} | u_{v\vec{k}} \rangle \approx \langle \psi_{c\vec{k}} | \vec{p} | \psi_{v\vec{k}} \rangle$ and therefore arrive

$$\left\langle \psi_{c\vec{k}+\vec{Q}} \middle| e^{i\vec{q}\cdot\vec{r}} \vec{p} \middle| \psi_{v\vec{k}} \right\rangle \approx \delta_{\vec{q}_\parallel,\vec{Q}} \langle \psi_{c\vec{k}} | \vec{p} | \psi_{v\vec{k}} \rangle, \tag{S23}$$

which can be regarded as the electric dipole approximation.

Combining Eqs.(S20-S23), we reach the closed form of the LMI matrix element,

$$\left\langle \Psi_{S\vec{Q}}^X \middle| \widehat{H}_I \middle| GS \right\rangle = \frac{E_g}{2i\hbar} \mathcal{A}_\ell(\vec{Q}) \left( \hat{\varepsilon} \cdot \vec{D}_{S\vec{Q}}^{X*} \right) e^{-i\omega t}. \tag{S24}$$

Considering Eq.(S11) and the OAM-light with the generic polarization written as $\hat{\varepsilon} = \alpha_R \hat{\varepsilon}_+ + \alpha_L \hat{\varepsilon}_-$ (with $|\alpha_R|^2 + |\alpha_L|^2 = 1$) in the basis of the $\sigma^\pm$-circular polarization, one can formulate the matrix element of LMI in Eq.(S24) for $\Psi_{S_\pm\vec{Q}}^X$ as

$$\left\langle \Psi_{S_\pm\vec{Q}}^X \middle| \widehat{H}_I \middle| GS \right\rangle = \frac{E_g \sqrt{\Omega} \, D_0^X}{2i\hbar} \mathcal{A}_\ell(\vec{Q})(\alpha_R \pm \alpha_L \, e^{-i2\varphi_{\vec{Q}}}) e^{-i\omega t}. \tag{S25}$$

**III-1. Rate of the optical transition induced by an OAM light (Derivation of Eq.4)**

Considering the exciting light beam with sufficiently broad band width, the total rate of the optical transition to excite the exciton state $\Psi^X_{S\vec{Q}}$ is evaluated by integrating Eq.(S19) over the angular frequency as given by

$$W^{\hat{\varepsilon}\ell}_{S\vec{Q}} = \int d\omega\, w^{\hat{\varepsilon}\ell}_{S\vec{Q}}(\omega) = \frac{\pi E_g^2}{2\hbar^4}\left|\mathcal{A}_\ell(\vec{Q})\left(\hat{\varepsilon}\cdot\vec{D}^{X*}_{S\vec{Q}}\right)\right|^2. \tag{S26}$$

From Eqs.(S19), (S25) and (S26), the total rate of the optical transition to excite the exciton state $\Psi^X_{S_\pm \vec{Q}}$ is derived as

$$W^{\hat{\varepsilon}\ell}_{S_\pm \vec{Q}} = \frac{\pi E_g^2 \Omega (D_0^X)^2}{2\hbar^4}\left|\mathcal{A}_\ell(\vec{Q})(\alpha_R \pm \alpha_L\, e^{-i2\varphi_{\vec{Q}}})\right|^2. \tag{S27}$$

### III-2. The spectral density function of optical transition (derivation of $g^\ell_S(E)$ in Eq.5)

To simulate the spectral profile of the optical lines induced by an OAM light, we specify the spectral density function, $g^\ell_S(E)$, that resolves the total transition rate in energy

$$W^{\hat{\varepsilon}\ell}_S = \sum_{\vec{Q}} W^{\hat{\varepsilon}\ell}_{S\vec{Q}} = \frac{\Omega}{4\pi^2}\int d^2\vec{Q}\, W^{\hat{\varepsilon}\ell}_{S\vec{Q}} = \int dE\, g^\ell_S(E). \tag{S28}$$

From Eq.(S27) and (S28), we can rewrite the total rate of the optical transition involving the exciton state $\Psi^X_{S_\pm \vec{Q}}$ as

$$W^{\hat{\varepsilon}\ell}_{S_\pm} = \frac{E_g^2 \Omega^2 (D_0^X)^2}{8\pi\hbar^4}\int_0^{2\pi} d\varphi_{\vec{Q}}\left|\alpha_R \pm \alpha_L\, e^{-i2\varphi_{\vec{Q}}}\right|^2 \int_0^\infty dQ\, Q|\mathcal{A}_\ell(\vec{Q})|^2 = \beta \int_0^\infty dQ\, Q|\mathcal{A}_\ell(Q)|^2, \tag{S29}$$

where we take the value of the angular integral given by,

$$\int_0^{2\pi} d\varphi_{\vec{Q}}\left|\alpha_R \pm \alpha_L\, e^{-i2\varphi_{\vec{Q}}}\right|^2 = 2\pi \tag{S30}$$

and $\beta \equiv E_g^2 \Omega^2 (D_0^X)^2 / 4\hbar^4$. To relate Eqs.(S28) and (S29), we need to use the explicit exciton band dispersion, $E(\vec{Q}) \equiv E^X_{S\vec{Q}}$, for the variable transform for the equations.

***The upper linear exciton band*** For the linear exciton upper band, $E(\vec{Q}) \equiv E^X_{S_+\vec{Q}} = 2\gamma Q$ is given by Eq.(S5), where $\gamma$ stands for the strength of the electron-hole exchange interaction in the valley exciton with $\vec{Q}$. Hereafter, let us set $E_0 = 0$ in Eq.(S5) for the simplicity of analysis. Accordingly, one can show the spectral density function for the *linear* band as

$$g_{S_+}^\ell(E) = \bar{P}_{S_+}^\ell(E)\rho_{S_+}(E) = \beta_\ell' \left[\frac{E}{(2\gamma)^2}\right] \left[\left(\frac{E^2}{4\gamma^2}\right)\left(\frac{W_0^2}{2}\right)\right]^{|\ell|} \exp\left[-\left(\frac{E^2}{4\gamma^2}\right)\left(\frac{W_0^2}{2}\right)\right] \quad (S31)$$

where $\beta_\ell' \equiv E_g^2 (D_0^X)^2 \pi A_0^2 W_0^4 / 2\hbar^4 |\ell|!$, $\rho_{S_+}(E) \propto E^1$ is the density of the exciton states with the linear dispersion, and $\bar{P}_{S_+}^\ell(E) = 2\pi\beta |\mathcal{A}_\ell(E/2\gamma)|^2/\Omega$ is defined as the averaged transition for the exciton states lying in the energy interval between $E$ and $E + dE$.

***The lower parabolic exciton band*** From Eq.(S6), we obtain the $E - Q$ relationship,

$$Q = \sqrt{2M_{\text{eff}}^X E}/\hbar \quad (S32)$$

for the parabolic exciton upper band. Accordingly, one can show the spectral density function for the *parabolic* band as

$$g_{S_-}^\ell(E) = \bar{P}_{S_-}^\ell(E)\rho_{S_-}(E) = \beta_\ell' \left[\frac{M_{\text{eff}}^X}{\hbar^2}\right] \left[\left(\frac{2M_{\text{eff}}^X E}{\hbar^2}\right)\left(\frac{W_0^2}{2}\right)\right]^{|\ell|} \exp\left[-\left(\frac{2M_{\text{eff}}^X E}{\hbar^2}\right)\left(\frac{W_0^2}{2}\right)\right], \quad (S33)$$

where $\bar{P}_{S_-}^\ell(E) = 2\pi\beta \left|\mathcal{A}_\ell\left(\sqrt{2M_{\text{eff}}^X E}/\hbar\right)\right|^2/\Omega$ and $\rho_{S_-}(E) \propto E^0$ is the density of the exciton states with the parabolic dispersion.

From Eqs.(S28-33), one can evaluate the averaged energy of the spectral density function,

$$g^\ell(E) \equiv g_{S_+}^\ell(E) + g_{S_-}^\ell(E), \quad (S34)$$

with respect to the exciting LG light with the OAM $\ell$, according to the definition

$$\langle E \rangle_\ell = \frac{\int dE\, E g^\ell(E)}{\int dE\, g^\ell(E)} = \frac{\int dE\, E g_{S_+}^\ell(E)}{\int dE\, g^\ell(E)} + \frac{\int dE\, E g_{S_-}^\ell(E)}{\int dE\, g^\ell(E)} = \langle E \rangle_\ell^{S_+} + \langle E \rangle_\ell^{S_-}, \quad (S35)$$

where $\langle E \rangle_\ell^{S_\pm} \equiv \int dE\, E g_{S_\pm}^\ell(E)/\int dE\, g^\ell(E)$.

In fact, $\langle E \rangle_\ell^{S_+}$ and $\langle E \rangle_\ell^{S_-}$ follows the different $\ell$-dependence because the distinctive band dispersion for $S_+$ and $S_-$ leads to the different $\ell$-dependences of the density of the exciton states and the averaged transition rate as functions of the energy E. For illustration, Fig.S7 plots the density of the exciton states, the averaged transition rate, and the spectral density function for the both linear and parabolic exciton bands as functions of the energy $E$.

**III-3. The OAM-light induced spectral shifts**

***The upper linear exciton band*** With the explicit expression of $g_{S_+}^\ell(E)$ given by Eq.(S31), one can show that the average spectral energy of the upper linear band is

$$\langle E \rangle_\ell^{S_+} \equiv \frac{\int dE\, E g_{S_+}^\ell(E)}{\int dE\, g^\ell(E)} = \left(\frac{\sqrt{2\pi}\gamma}{W_0}\right) \frac{1}{2^{|\ell|+1}} \frac{(2|\ell|+1)!!}{|\ell|!}, \quad (S36)$$

and the spectral energy shift

$$\Delta E_\ell^{S_+} \equiv \langle E \rangle_\ell^{S_+} - \langle E \rangle_{\ell=0}^{S_+} = \left(\frac{\sqrt{2\pi}\gamma}{W_0}\right) \left[\frac{1}{2^{|\ell|+1}} \frac{(2|\ell|+1)!!}{|\ell|!} - \frac{1}{2}\right]. \quad (S37)$$

Eq.(S37) manifests the *non-linear* $\ell$-dependence of the spectral energy shift for a *linear* exciton band under the $\ell$-OAM light excitation.

***The lower parabolic exciton band*** With the explicit expression of $g_{S_-}^\ell(E)$ given by Eq.(S33), the average spectral energy of the parabolic band is shown as

$$\langle E \rangle_\ell^{S_-} \equiv \frac{\int dE\, E g_{S_-}^\ell(E)}{\int dE\, g^\ell(E)} = \left(\frac{\hbar^2}{2M_{\text{eff}}^X W_0^2}\right) [|\ell| + 1], \quad (S38)$$

and the OAM-light-induced spectral energy shift is

$$\Delta E_\ell^{S_-} \equiv \langle E \rangle_\ell^{S_-} - \langle E \rangle_{\ell=0}^{S_-} = \left(\frac{\hbar^2}{2M_{\text{eff}}^X W_0^2}\right) |\ell| \propto |\ell|. \quad (S39)$$

In contrast to the linear exciton band, Eq.(S38) predicts the *linear* $\ell$-dependence of the spectral energy shift for a *parabolic* exciton band under the $\ell$-OAM light excitation. The analysis here advices us to infer the types of exciton band dispersion, in the linearity or parabolicity, by examining the $\ell$-dependence of the spectral energy shift induced by the OAM light. As the non-linear $\ell$-dependence of the spectral energy shift was apparently observed and falls in the similar energy scale of the predicted energy of the linear upper exciton band in our experiment, our spectroscopic measurement provides the evidence of the existence of the predicted lightlike exciton band in TMD-MLs [2].

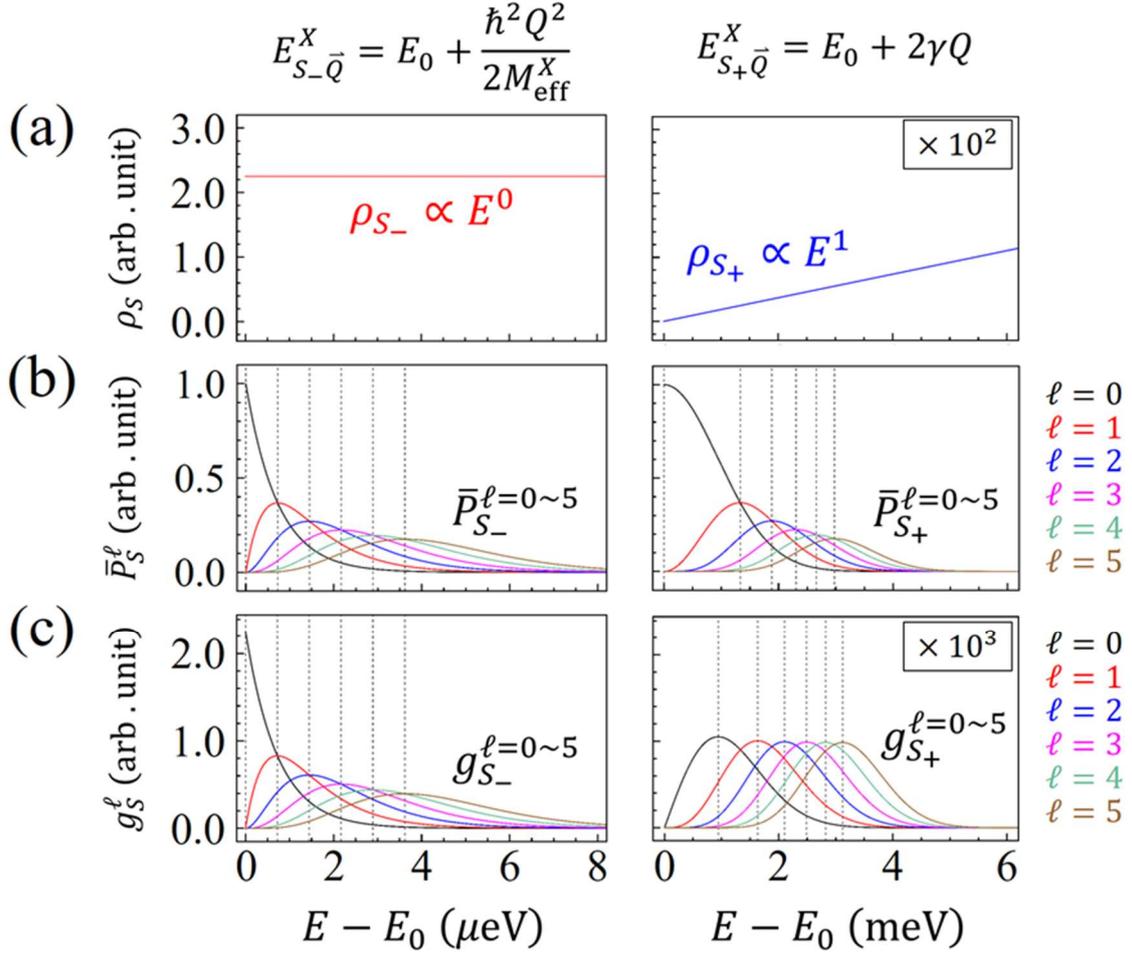

**Figure S7.** **(a)** The density of exciton states, **(b)** the average transition rate, and **(c)** the spectral function as functions of energy $E$ for both the parabolic lower $S_-$ (left panels) and linear upper $S_+$ (right panels) exciton bands of a ML-MoS$_2$ under the photo-excitation of twisted beams with $\ell = 0, 1, \ldots 5$.